\documentclass{jaa}

\usepackage{natbib}
\usepackage{graphicx,amsmath,amsfonts}

\begin{document}\sloppy

\title{New parameterizations of generalized Chaplygin gas model constrained at background and perturbation levels}


\author{S. F. Salahedin\textsuperscript{1}, M. Malekjani\textsuperscript{2}, K. Y. Roobiat\textsuperscript{1} and R. Pazhouhesh\textsuperscript{1,*}}

\affilOne{\textsuperscript{1}Department of Physics, Faculty of Sciences, University of Birjand , Birjand, Iran.\\}

\affilTwo{\textsuperscript{2}Department of Physics, Bu-Ali Sina University, Hamedan 65178, Iran.}


\twocolumn[{

\maketitle

\corres{rpazhouhesh@birjand.ac.ir}

\msinfo{1 January 2021}{1 January 2021}

\begin{abstract}
We study the main cosmological properties of the Generalized Chaplygin Gas (GCG) dark energy model at the background
and perturbation levels. By using the latest cosmological data in both the background and perturbation levels, we
implement a joint likelihood analysis to constrain the cosmological parameters of the model. Using the available
expansion and growth rate data, we place constraints on the free parameters of the GCG model based on the statistical
Markov chain Monte Carlo method. Then, the best-fit values of cosmological parameters and those of confidence regions
are found. We obtain the best-fit value of the current expansion rate of the universe in the GCG model and show that it
is in good agreement with the $\Lambda$CDM model. Moreover, the growth rate of matter perturbations is investigated in
the context of a unified GCG model. It is shown that in this model, the dark energy component, like the $\Lambda$ sector
in the $\Lambda$CDM model, can suppress the amplitude of matter perturbations. We show that the growth rate of
perturbations in GCG parametrization is consistent with cluster-scale observations similar to the case of the
concordance $\Lambda$CDM model. Our results show that the tension on $\sigma_{8}$ appeared in concordance model can be
alleviated in GCG cosmology.
\end{abstract}


}]



\doinum{12.3456/s78910-011-012-3}
\artcitid{\#\#\#\#}
\volnum{000}
\year{0000}
\pgrange{1--}
\setcounter{page}{1}
\lp{1}

\section{Introduction}

Several evidences from independent cosmic observations, including the rotation curve of spiral galaxies (\citealt{Persic1996}), large scale structure formation in the Universe (\citealt{Primack1996}), and dynamics of galaxy clusters (\citealt{Frenk1996}), indicate that in the cosmic matter budget, there is a roughly six times more cold dark matter (CDM) than what can be afforded by the baryonic matter making up $\sim 30\%$ of critical density (\citealt{Riess1998}). Besides this clustering dark component, the cosmological observations, such as supernovae Type Ia (SNIa) (\citealt{Riess1998}; \citealt{Kowalski2008}), Large-Scale Structure (LSS) by Sloan Digital Sky Survey (SDSS) (\citealt{Tegmark2004}), Cosmic Microwave Background (CMB) fluctuations (\citealt{Jarosik2011}), Baryonic galaxy clustering, and Acoustic Oscillation (BAO) (\citealt{Percival2010}; \citealt{Blake2011}), represent the presence of the so-called dark energy- an exotic fluid with enough negative pressure that leads to the late-time accelerated expansion of the universe.

The physical properties, origin, and nature of dark matter and dark energy are still unknown although many efforts have been made to study them. It has been proposed that a complete theory of quantum gravity has a pivotal role in understanding the nature of dark energy. It may be asked that, can a simple model be presented, in which a single dark fluid acts as both dark matter and dark energy (\citealt{Tavares2003})? Dark matter and dark energy can be unified by dark fluid with a barotropic Equation of State (EoS), which can then explain both the accelerated expansion at late times and decelerated expansion at earlier times. This means that at high redshifts, the barotropic EoS parameter of dark fluid behaves as the dark matter EoS parameter ($w_{m}\sim 0$), and at low redshifts, acts like the dark energy EoS parameter ($w_{de}<-\frac{1}{3}$). This duality is very astonishing and the coincidence problem of standard $\Lambda$CDM model can be solved for such a fluid (\citealt{Sahni2005}). The so-called Generalized Chaplygin Gas (GCG) is a special type of dark fluid, which unifies dark matter and dark energy (\citealt{Kamenshchik2001}; \citealt{Bento2002}).

The present study follows the line of the studies mentioned above, and attempts to investigate the GCG model at the background and perturbation levels. This paper is organized as follows. In section II, we introduce the GCG parameterization to investigate the redshift evolution of the main cosmological parameters in this model. The most recent cosmological data, including data from SNIa, CMB, Big Bang Nucleosynthesis (BBN), BAO, and Hubble expansion rate, are used in the Markov Chain Monte Carlo (MCMC) algorithm too constrain the GCG free parameters. In Section III, the matter perturbation growth in the GCG model is investigated. We place some constraints on the model parameters at the perturbation level by performing a joint likelihood analysis and using the perturbation growth rate data. In Sections IV, we present the results, discussion and finally we conclude this paper in section V.

\section{Generalized Chaplygin Gas model}
An interesting dark energy model, called the Chaplygin Gas (CG), was proposed by (\citealt{Kamenshchik2001}). The EoS of this model is given as follows:
\begin{equation}
     P_{CG}=-\frac{A}{\rho_{CG}},
\label{eq1}
\end{equation}
where A denotes a positive constant. The substitution of Eq. (\ref{eq1}) in the equation of energy conservation yields the energy density of CG as follows
\begin{equation}
     \rho_{CG}=\rho_{GC,0}\left[A_{c}+\frac{1-A_{c}}{a^6}\right]^{1 / 2},
\label{eq2}
\end{equation}
where $\rho_{GC,0}$ is the current energy density of CG and $A_{c}\equiv\frac{A}{\rho_{GC,0}}$. We can explicitly see that the CG model behaves like dark matter, $\rho_{GC}\propto a^{-3}$, at earlier times ($a<<1$), and mimics the cosmological constant at the present time ($a=1$). Also at present time, the CG model can recover both dark matter and cosmological constant energy densities by putting $A_c=0$ and $A_c=1$, respectively. For the interval range $0<A_c<1$, the energy density of CG can be considered as a combination of dark matter and dark energy. Note that for $0<A_c<1$, the CG model model can behave like dark energy with time varying energy density different from constant cosmological constant $\Lambda$. The CG model can be generalized to Generalized Chaplygin Gas (GCG) model by including a new parameter in the EoS (\citealt{Bento2002}) as
\begin{equation}
     P_{GCG}=-\frac{A}{\rho_{GCG}^\alpha},
\label{eq3}
\end{equation}
where $A$ and $\alpha$ are the parameters of the model. A limiting case CG model (\citealt{Kamenshchik2001}) can be recovered by $\alpha=1$ and also the cosmological constant $\Lambda$ is achieved by $\alpha=0$. Using the continuity equation, the energy density of GCG model is obtained as
\begin{equation}
     \rho_{GCG}=\rho_{GCG,0}\left[A_{s}+\frac{1-A_{s}}{a^{3(1+\alpha)}}\right]^{1 / (1+\alpha)},
\label{eq4}
\end{equation}
where $a$ is the scale factor, $A_s\equiv\frac{A}{\rho_{GCG}^{1+\alpha}}$, and $\rho_{GCG,0}$ is the current energy density of the GCG model. We observe that the energy density of GCG fluid changes from $\rho_{GCG}\propto a^{-3}$ at early times ($a<<1$) to $\rho_{GCG}=\rho_{GCG,0}$ at the present time. This behavior of GCG model represents the matter dominated Universe with decelerated expansion at earlier times and dark energy dominated Universe with accelerated expansion at the current time. It is also interesting to study the feature of GCG model according to model parameter $A_s$. In the limiting case $A_s = 0$, we have $\rho_{GCG}\propto a^{3}$ indicating purely matter fluid and in the limiting case $A_s = 1$, we have $\rho_{GCG}\propto \rho_{GCG,0}$ representing purely cosmological constant. Note that for other values, we have neither purely matter nor cosmological constant. By combining Eqs. (\ref{eq3}) and (\ref{eq4}) and using the continuity equation, the EoS parameter of GCG can be obtained as follows:
\begin{equation}
     w_{GCG}=-\frac{A_{s} a^{3(1+\alpha)}}{1-A_{s}+A_{s} a^{3(1+\alpha)}}.
\label{eq5}
\end{equation}
It is clear from Eq.(\ref{eq5}) that at early times ($a\to 0$), the EoS tends to zero ($w_{GCG}\to 0$) and at far future ($a\to\infty$), $w_{GCG}\to -1$. we conclude that both early matter dominated and future $\Lambda$-dominated phases of the Universe can be recovered in GCG model. So the EoS parameter of GCG model is limited to the interval $-1<w_{GCG}<0$, representing the time varying EoS parameter of dynamical dark energy scenario in quintessence regime. Additionally, Eq.(\ref{eq5}) indicates that we have $ w_{GCG} < -1$ for $ A_s > 1 $ which means that the EoS parameter of GCG can cross the phantom divide and varies in phantom regime ($w_{GCG}<-1$). Since the GCG model unifies the dark matter and dark energy, this model can be decomposed into two components as follows:
\begin{equation}
     \rho_{GCG}=\rho_{de}+\rho_{dm},
\label{eq6}
\end{equation}
where $\rho_{de}$ and $\rho_{dm}$ are the energy densities of dark energy and dark matter, respectively. Note that since the dark matter component is pressureless, we have $P_{GCG}=P_{de}$. We know that the evolution of energy density of dark matter as
\begin{equation}
     \rho_{dm}=\rho_{dm,0} a^{-3},
\label{eq7}
\end{equation}
so the evolution of dark energy density in the GCG model is obtained as
\begin{equation}
\begin{aligned}
     \rho_{de}&=\rho_{GCG}-\rho_{dm} \\
     & =\rho_{GCG,0}\left[A_s+(1-A_s)a^{-3(1+\alpha)}\right]^{1 / (1+\alpha)} \\
     & -\rho_{dm,0} a^{-3}.
\end{aligned}
\label{eq8}
\end{equation}
On the other hand, we know that the total energy of the Universe in GCG model is the combination of GCG fluid and baryonic matter as $\rho_t=\rho_{GCG}+\rho_b$. So in this context and for a flat FRW Universe, the Freidmann equation reads	

\begin{equation}
\begin{aligned}
     H^2&=\frac{8\pi G}{3}\rho_{t} \\
     & =\frac{8\pi G}{3}\bigg(\rho_{GCG,0}\left[A_s+(1-A_s)a^{-3(1+\alpha)}\right]^{1 / (1+\alpha)} \\
     & +\rho_{b,0} a^{-3}\bigg),
\end{aligned}
\label{eq9}
\end{equation}
where $\rho_{b,0}$ is the current baryonic matter density. Inserting the following dimensionless parameters,
\begin{equation}
     \rho_{GCG,0}=\frac{3 {H_{0}}^2}{8 \pi G}\Omega_{GCG},\:\:\: \rho_{b,0}=\frac{3 {H_{0}}^2}{8 \pi G}\Omega_{b},\:\:\: \Omega_{GCG}+\Omega_{b}=1,
\label{eq10}
\end{equation}
into Eq.(\ref{eq9}), the Hubble parameter can be written as
\begin{equation}
\begin{aligned}
     H^2 & ={H_{0}}^2 E(a)^2 \\
     & ={H_{0}}^2\bigg((1-\Omega_b)\left[A_s+(1-A_s)a^{-3(1+\alpha)}\right]^{1 / 1+\alpha}\\
     &+\Omega_b a^{-3}\bigg),
\end{aligned}
\label{eq11}
\end{equation}
where $E(a)$ is the normalized Hubble parameter, and is given by
\begin{equation}
\begin{aligned}
     E(a)^2=&\bigg((1-\Omega_b)\left[A_s+(1-A_s)a^{-3(1+\alpha)}\right]^{1 / 1+\alpha}\\
     &+\Omega_b a^{-3}\bigg)^{1 / 2}.
\end{aligned}
\label{eq12}
\end{equation}
Now, $\rho_{dm}=\left(\frac{w_{de}(a)-w_{GCG}(a)}{w_{de}(a)}\right)\rho_{GCG}$ and $\rho_{de}=\frac{w(a)}{w_{de}(a)}\rho_{GCG}$ can be obtained from Eq. (\ref{eq6}). Hence, the dimensionless energy densities of dark matter and dark energy can be obtained as $\Omega_{dm}=\left(\frac{w_{de}(a)-w_{GCG}(a)}{w_{de}(a)}\right)\Omega_{GCG}$ and $\Omega_{de}=\frac{w(a)}{w_{de}(a)}\Omega_{GCG}$. Finally, we have
\begin{equation}
\begin{aligned}
     \Omega_{dm}=&\left(w_{de}(a)-w_{GCG}(a)\right)\times \\
     &\frac{\left(1-\Omega_b-\Omega_r\right)\left[A_s+(1-A_s)a^{-3(1+\alpha)}\right]}{w_{de}(a)E^2(a)},
\end{aligned}
\label{eq13}
\end{equation}
\begin{equation}
     \Omega_{de}=\frac{w_{GCG}(a)\left(1-\Omega_b-\Omega_r\right)\left[A_s+(1-A_s)a^{-3(1+\alpha)}\right]}{w_{de}(a)E^2(a)}.
\label{eq14}
\end{equation}

\section{Cosmological constraints from geometrical observations}
\label{s3}
According to the background expansion data from binned sample of Type Ia Supernovae (SNIa)(\citealt{Betoule2014}), WMAP Planck data for the position of CMB acoustic peak (\citealt{Hu1996}), baryon acoustic oscillation (BAO) (\citealt{Blake2011}), BBN (\citealt{Serra2009}), and Hubble data ($H(z)$) extracted from cosmic chronometers (\citealt{Moresco2012}), we perform a statistical Markov Chain Monte Carlo (MCMC) analysis for the GCG model. For more details of the MCMC method, we refer the reader to (\citealt{Mehrabi2015a}). It is worth mentioning that the following sets of data are used: 580 distinct points for SNIa observation in Union2.1 catalog, 26 points for Hubble data, BAO data including 6 distinct measurements of the baryon acoustic scale, BBN data points with $\Omega_b$, and the WMAP data. In this joint likelihood analysis, the product of the individual likelihood for each experiment is considered as the total likelihood function as
\begin{equation}
     \mathcal{L}_{t}(P)=\mathcal{L}_{SN} \times \mathcal{L}_{BAO} \times \mathcal{L}_{CMB} \times \mathcal{L}_{BBN} \times \mathcal{L}_{H}.
\label{eq15}
\end{equation}
Therefore, the total chi-square ($\chi_{t}^{2}$) is
\begin{equation}
     \chi_{t}^{2}=\chi_{SN}^{2}+\chi_{BAO}^{2}+\chi_{CMB}^{2}+\chi_{BBN}^{2}+\chi_{H}^{2}.
\label{eq16}
\end{equation}
The free parameters that should be constrained in the MCMC algorithm are ($\Omega_{b,0}$, $\Omega_{dm,0}$, $H_{0}$) in standard $\Lambda$CDM cosmology and are ($\Omega_{b,0}$, $A_s$, $H_0$, $\alpha$) in GCG cosmology.
We know that a model with a lower value of $\chi^2$ is in a better agreement with the observational data, if the numbers of free parameters of the two models are equal. Obviously, this analysis is no longer valid if we compare different models with different numbers of free parameters. Therefore, we have to employ another statistical test (the so-called Akaike Information Criteria ($AIC$) (\citealt{Akaike1974})) to compare GCG and $\Lambda$CDM models. Notice that in the $\Lambda$CDM and GCG models, there are three and four free parameters, respectively.

\begin{figure*}
\centering
\begin{tabular}{@{}c@{}}
\resizebox{0.75\textwidth}{!}{%
\includegraphics{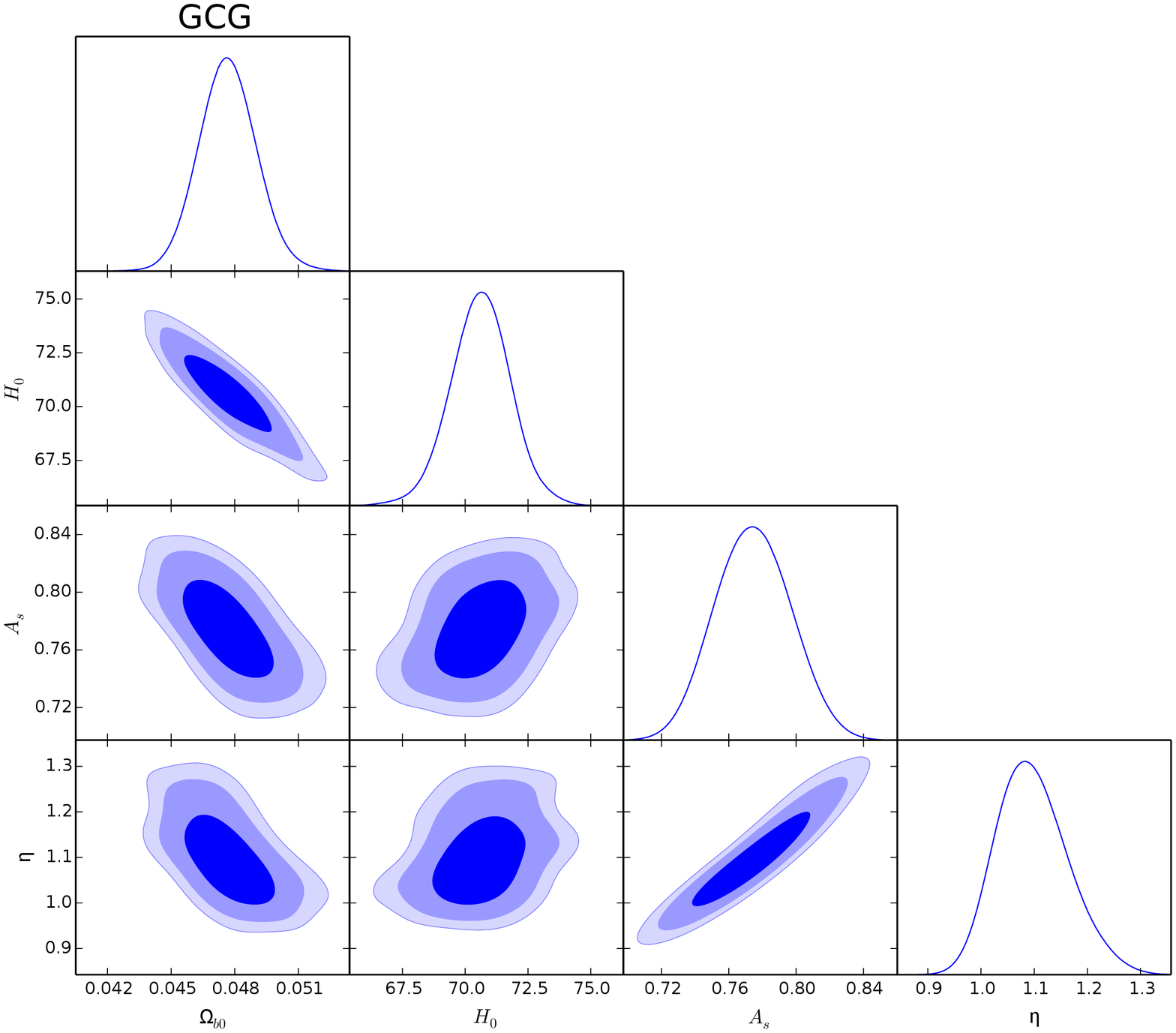}
}\\
\resizebox{0.75\textwidth}{!}{%
\includegraphics{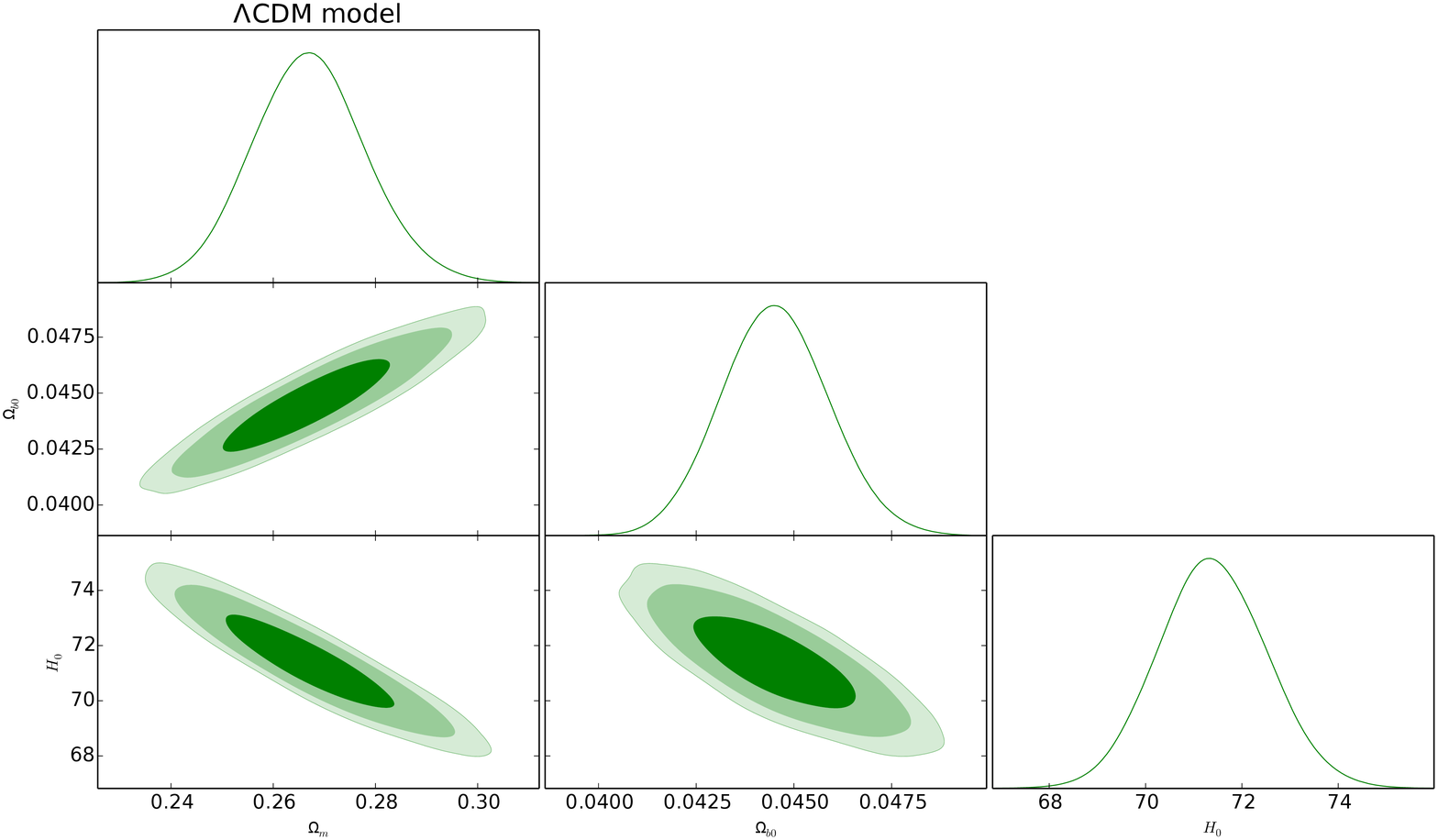}
}\\
\end{tabular}
\caption{: Confidence levels 1$\sigma$, 2$\sigma$ and 3$\sigma$ and maximum likelihood function for different cosmological parameters in GCG and $\Lambda$CDM models. }
\label{fig1}
\end{figure*}

Our numerical results are presented in Table (\ref{tab1}). According to Table (\ref{tab1}), for the joint data sets $H(z)$+BAO+BBN+CMB+SNIa (Union2.1), we obtain $\chi_{min}^2 = 592.6$ in the GCG model and $\chi_{min}^2=571.9$ in the $\Lambda$CDM model. For eliminating the effect of extra parameters, we use the relation $AIC = \chi_{min}^2 + 2k$, in which $k$ is the number of free parameters (\citealt{Akaike1974}). The $AIC$ values for the two models are $AIC_{\Lambda CDM} = 577.9$ and $AIC_{GCG} = 600.6$; the large value of $\Delta AIC = 22.7$ shows the better agreement of the standard $\Lambda$CDM model with the above cosmological observations (for a detailed discussion on $\Delta AIC$, see (\citealt{Burnham2002})). Based on the statistical likelihood analysis, the standard $\Lambda$CDM model is still better than the GCG model. In previous works on the dynamical dark energy, the same results have been achieved (for example, see (\citealt{Mehrabi2015a}; \citealt{Rezaei2017}; \citealt{Rezaei2019}; \citealt{Malekjani2018})). To the best of our knowledge, no dynamical dark energy with a time-varying EoS parameter and an $AIC$ less than that of the standard $\Lambda$CDM model has been reported. The best-fit values of the cosmological parameters are also shown in Table (\ref{tab1}). Moreover, Figure (\ref{fig1}) shows $1\sigma$, $2\sigma$, and $3\sigma$ confidence levels of cosmological parameters. We see that the Hubble constant $H_0$ obtained in the GCG model is well consistent with that in the concordance $\Lambda$CDM model with $1\sigma$ uncertainty (see also Table (\ref{tab1})). This indicates that the expansion rate of the Universe in the GCG model is comparable with that of the standard $\Lambda$CDM universe.

\begin{figure}[!t]
\centering
\includegraphics[width=.8\columnwidth]{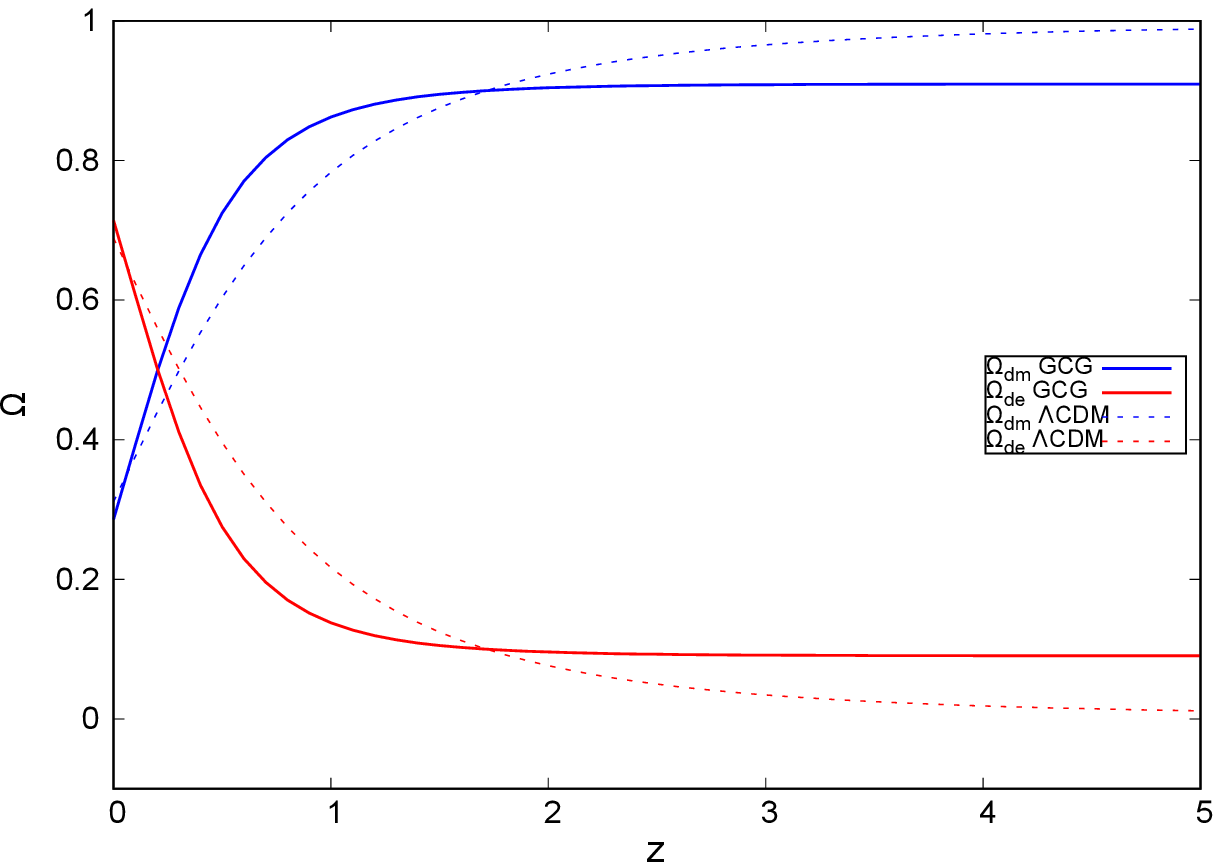}

\caption{: The Evolution of dimensionless energy density of non-relativistic matter and dark energy components in terms of redshift for GCG and $\Lambda$CDM models using the best-fit values from Table (\ref{tab1}).}
\label{fig2}
\end{figure}

Figure (\ref{fig2}) shows the evolution of dimensionless energy densities for dark matter and dark energy in the context of the GCG model with the best-fit values of the cosmological parameters shown in Table (\ref{tab1}). We observe that in the context of GCG model, dark energy dominates at redshift $z_t=0.203$. This means that before $z_t=0.203$ until matter-radiation equality epoch, the Universe is dominated by dark matter and after $z_t=0.203$ the expansion of the Universe enters to accelerated phase by dominating the dark energy component. In the case of $\Lambda$CDM model, the transition redshift from early decelerated expansion to current accelerated expansion is obtained as $z_t=0.303$. Hence, we conclude that the transition phase takes place later in GCG cosmology compare to $\Lambda$CDM cosmology.

\begin{table*}
 \centering
 \tabularfont
 \caption{Best fit values of the cosmological parameters resulted from MCMC analysis by using the observational data in background level for GCG and $\Lambda$CDM models.}
\begin{tabular} { l  c c c c c c c  c c}
\hline		
 Parameters &GCG &$\Lambda$CDM\\
\hline
{$\Omega_b$} &
0.0477 $ \begin{array}{c}{+0.0014+0.0027+0.0038} \\ {-0.0014-0.0026-0.0034}\end{array}$
& 0.0445$ \begin{array}{c}{+0.0014+0.0028+0.0036} \\ {-0.0014-0.0027-0.0034}\end{array}$
\\
 {$H_0$} &
 70.6$ \begin{array}{c}{+1.2+2.4+3.3} \\ {-1.2-2.4-3.5}\end{array} $
 &71.4$ \begin{array}{c}{+1.1+2.3+3.0} \\ {-1.1-2.2-2.9}\end{array}$
\\
{$A_s$}  &
 0.774$ \begin{array}{c}{+0.022+0.044+0.054} \\ {-0.022-0.043-0.051}\end{array}$
 & -
 \\
{$\eta$}  &
1.096$\begin{array}{c}{+0.059+0.14+0.18} \\ {-0.074-0.12-0.15}\end{array}$
& -
\\
{$\chi_{min}^{2}$ }&
592.6 &571.9
\\
\hline
\end{tabular}\label{tab1}
\end{table*}

\begin{figure}[!t]
\centering
\begin{tabular}{@{}c@{}}
\includegraphics[width=.8\columnwidth]{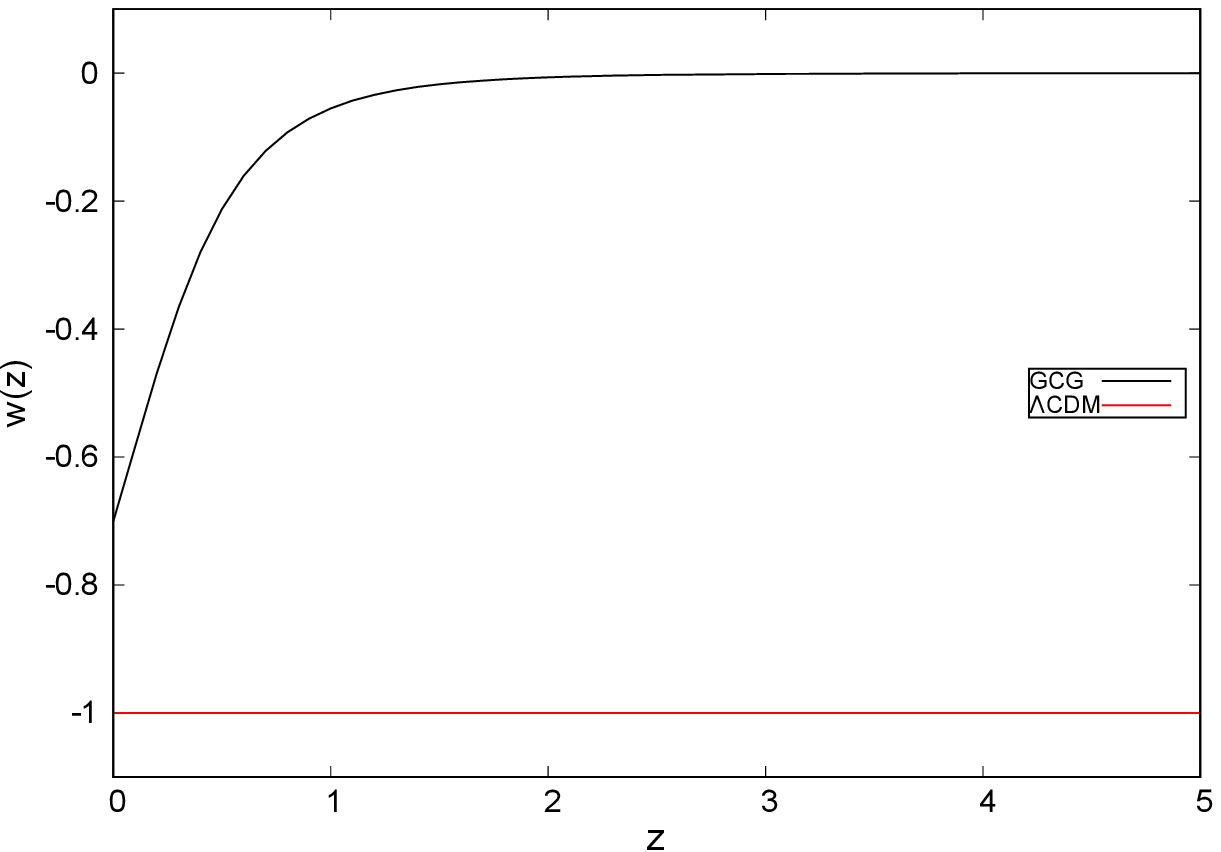}
\\
\includegraphics[width=.8\columnwidth]{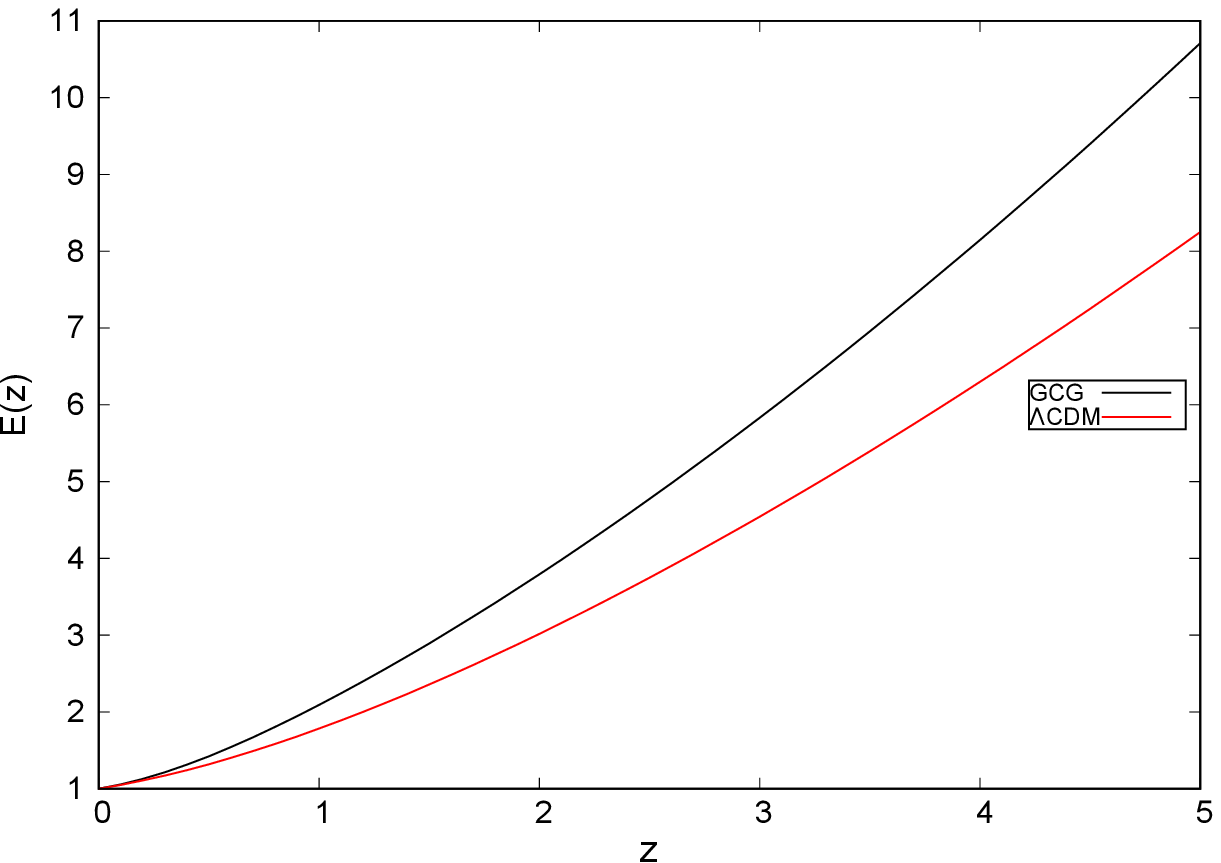}
\\
\end{tabular}
\caption{: Redshift evolution of EoS (top panel) and Hubble parameter (bottom panel) for GCG and $\Lambda$CDM models.}
\label{fig3}
\end{figure}

Figure (\ref{fig3}) shows the evolution of the EoS parameter, $w_{de}$, and the dimensionless Hubble parameter, $E=H/H_0$, as a function of redshift, $z$ for both GCG and $\Lambda$CDM models. We can see that the EoS of the GCG model tends to zero at high redshifts, representing a pressure-less matter domination ($w = 0$). However, EoS decreases to values smaller than $-\frac{1}{3}$ as the redshift decreases. this behavior is required for GCG to act as dark energy at current epoch. Moreover, we see that $E$ is smaller in the $\Lambda$CDM model than in the GCG model at high redshifts, while they coincide at low redshifts. Therefore, the GCG model can reproduce the current Universe expansion rate in the same way as $\Lambda$CDM.

\section{Perturbations growth}
\label{s4}
We now investigate the GCG model in cluster scale, where the perturbations of matter is important. Since most of the dynamical dark energy models have the same behavior with $\Lambda$CDM cosmology at the expansion level, studying them in perturbation level may be considered as a possible tool to distinguish between them and standard cosmology. Therefore, the information come from Large scale Structure (LSS) formation in the Universe, provides us a powerful measurement to compare different dark energy models with standard model. Here, we first describe the growth of matter perturbation in the GCG model. Then GCG model is examined in cluster scale, using the observational growth rate data at the perturbation level. The general linear perturbation theory for unified dark matter and dark energy scenarios is studied in (\citealt{Davari2018}). In the linear perturbation theory, the perturbed FRW metric is given by
\begin{equation}
     ds^2=a^2(\tau)\left[-(1+2\psi)d\tau^2+(1-2\varphi)\delta_{ij}dx^idx^j\right],
\label{eq17}
\end{equation}
where $x$ is the spatial coordinate, and $\psi$ and $\varphi$ are small perturbations around isotropic and homogeneous Universe. The perturbed metric in the absence of anisotropic stress is given by
\begin{equation}
     g_{\mu\nu}=\left(\begin{array}{ccccc}{-(1+2\varphi)} & {0} & 0 & 0 \\ {0} & {(1-2\varphi)} & 0 & 0\\ 0 & 0 & {(1-2\varphi)} & 0 \\ 0 & 0 & 0 & {(1-2\varphi)} \end{array}\right).
\label{eq18}
\end{equation}
The perturbed metric in Eq. (\ref{eq18}) can be expanded as homogeneous metric $\bar{g}_{\mu\nu}$ plus perturbation term $h_{\mu\nu}$, where a time-time component of metric is $\bar{g}_{00}=-a^2$ and spatial part is $\bar{g}_{ij}=a^2\delta_{ij}$. Also the time-time component of perturbation term is $h_{00}=-2a^2\varphi$, and spatial term is $h_{ij}=-2a^2\varphi\delta_{ij}$ (\citealt{Marcondes2016}; \citealt{Davari2018}). From Einstein field equation, we know that the perturbation in metric is related to perturbation in energy-momentum tensor. At background level, the energy-momentum tensor for an isotropic and homogeneous Universe is given by 
\begin{equation}
     \bar{T}_{\mu\nu}=\bar{P} \bar{g}_{\mu\nu}(\bar{\rho}+\bar{P})\bar{u}_{\mu}\bar{u}_{\nu},
\label{eq19}
\end{equation}
where $u_{\mu}$ is the four-velocity, and the bars denote unperturbed quantities. In cluster scales, the total energy-momentum tensor can be decomposed as $T_{\mu\nu}=\bar{T}_{\mu\nu}+\delta T_{\mu\nu}$, where  $\delta T_{\mu\nu}$ represents the perturbation of energy- momentum tensor and can be obtained as (\citealt{Davari2018})
\begin{equation}
     \delta T_{\mu\nu}=(\delta_{\rho}+\delta_{P})\bar{u}_{\mu}\bar{u}_{\nu}+(\bar{\rho}+\bar{P})(\delta u_{\mu}\bar{u}_{\nu}+\bar{u}_{\mu}\delta u_{\nu})-\delta_{P}\delta_{\mu\nu}.
\label{eq20}
\end{equation}
Based on the above equations, the evolution of matter-density perturbation $\delta\equiv\frac{\delta\rho}{\bar{\rho}}$  and divergence of velocity of perturbations $\theta\equiv a^{-1}ik^j\delta u_j$ are given by (\citealt{Marcondes2016})
\begin{equation}
\begin{aligned}
&\dot{\delta}+\left[3\mathcal{H}w_{de}+\frac{\bar{Q_0}}{\rho}\right]\delta-(1+w_{de})\times \\
&(\theta-3\dot{\phi})=\frac{\delta Q_0}{\rho},\\
&\dot{\theta}+\left[\mathcal{H}(1-3w_{de})-\frac{\bar{Q_0}}{\rho}+\frac{\dot{w_{de}}}{1+w_{de}}\right]\theta-k^2\varphi \\
&-\frac{c_s^2 k^2}{1+w_{de}}\delta =\frac{ik^i\delta Q_{i}}{\bar{\rho}(1+w_{de})},
\end{aligned}
\label{eq21}
\end{equation}
where the over-dot denotes the derivative with respect to conformal time, $k^i$ are the components of the wave-vector in Fourier space, $\delta Q_i$ is the perturbations of energy-momentum exchange in the perturbed conservation equations, and $\mathcal{H}$ is the conformal Hubble parameter. Note that $\bar{Q_0}$ in Eq.(\ref{eq21}) is the energy exchange between dark matter and dark energy components. Moreover, for the homogeneous and isotropic universe, spatial components of $\bar{Q_0}$ are zero at the background level. Beside the equations for the evolution of $\delta$ and $\theta$, the perturbed Poisson equation for a perturbed fluid is given by
\begin{equation}
     (1+\frac{3\mathcal{H}^2}{k^2})k^2\phi=-3\mathcal{H}\dot{\phi}-4\pi Ga^2(\rho_m \delta_m),
\label{eq22}
\end{equation}
where $\delta_m$ denotes the density perturbation of pressure-less matter (baryons+dark matter). Note that the growth of perturbations whose their wavelengths much smaller than the horizon ($k \gg \mathcal{H}$) is of interest. So we assume the large size of sound horizon of dark energy and then we expect that dark energy cannot cluster in sub-horizon scales (\citealt{Duniya2013}). Therefore, the Pseudo-Newtonian cosmology can be used and the time variation of the gravitational potential can be neglected. Hence, in Eq. (\ref{eq22}), the second left-hand side term and the right-hand side term proportional to $\dot{\phi}$ can be ignored. Consequently, the perturbed Poisson equation reduces to
\begin{equation}
     k^2\phi=-4\pi Ga^2\rho_m\delta_m=-\frac{3}{2}\mathcal{H}^2\Omega_m\delta_m,
\label{eq23}
\end{equation}
where $\Omega_m$  is the sum of the dimensionless density parameter of dark matter and baryons (i.e., $\Omega_{dm}+\Omega_b$). Combining Eqs. (\ref{eq21}) and (\ref{eq23}), we have
\begin{eqnarray}\nonumber
&\dot{\delta}_m+3\mathcal{H}\xi\frac{\Omega_{de}}{\Omega_m}\delta_m+\theta_m=0,\\
&\dot{\theta}_m+\mathcal{H}(1+3\xi\frac{\Omega_{de}}{\Omega_m}\theta_m+\frac{3}{2}\mathcal{H}^2\Omega_m\delta_m=0.
\label{eq24}
\end{eqnarray}
After omitting $\theta_m$ in Eq. (\ref{eq24}) and transforming the variables from conformal time to physical time, we obtain
\begin{equation}
\begin{aligned}
     &\frac{d^2\delta_m}{dt^2}+2\left(H+3\xi\frac{\Omega_{de}}{\Omega_m}\right)\dot{\delta}_m-\frac{3}{2}H^2\bigg[\Omega_m-2\xi\frac{\Omega_{de}}{\Omega_m}\times \\
     &(1+\frac{\dot{H}}{aH^2}+3\xi\frac{\Omega_{de}}{\Omega_m}-\frac{\dot{\Omega}_{de}}{H\Omega_m\Omega_{de}})\bigg]\delta_m=0.
\end{aligned}
\label{eq25}
\end{equation}
Replacing the time derivative by the derivative with respect to the scaling factor, the following equation for the evolution of matter density contrast in sub-Hubble scale is obtained.
\begin{equation}
     \delta_m^{''}+A_m\delta_m^{'}+B_m\delta_m=S_m,
\label{eq26}
\end{equation}
where coefficients $A_m$, $B_m$ and $S_m$ are defined as
\begin{eqnarray}\nonumber
&A_m=\frac{3}{a}+\frac{H^{'}}{H}+\frac{6\xi}{a}\frac{\Omega_{de}}{\Omega_m},\\
\nonumber
&B_m=-\frac{3}{2a^2}\left[-2\xi\frac{\Omega_{de}}{\Omega_m}\left(1+\frac{H^{'}}{H}+3\xi\frac{\Omega_{de}}{\Omega_m}-\frac{\Omega_m^{'}}{\Omega_m\Omega_{de}}\right)\right],\\
&S_m=-\frac{3}{2a^2}\Omega_m\delta_m.
\label{eq27}
\end{eqnarray}
Note that for $\xi=0$, the standard equation of the matter perturbation evolution is recovered. By numerically solving Eq. (\ref{eq26}), we can obtain the growth of matter perturbation in dark energy cosmology. Based on the initial conditions, the initial scale factor $a_i=0.0005 $ $(z_i=2000)$ (\citealt{Batista2013}; \citealt{Mehrabi2015b}) implies being sufficiently deep at early matter-dominated era. To guarantee the linear regime of perturbation $(\delta_m<1)$ up to present time, we adopt the initial value of density contrast as $\delta_{mi}=8\times10^{-5}$ (\citealt{Batista2013}; \citealt{Mehrabi2015b}). Now, we use the background cosmological parameters from the best-fit values in Table (\ref{tab1}). Moreover, we consider the upper bound for interaction parameter ( $\xi\sim4\times10^{-4}$). After obtaining the matter perturbation $\delta_m(z)$, the evolution of the growth rate function, $f=\frac{d\ln\delta_m}{d\ln a}$, and the mass variance of matter perturbations, $\sigma_8$, can be calculated. The variance of perturbations at redshift $z$ reads $\sigma_8(z)=D(z)\sigma_8(z=0)$, where $D(z)=\frac{\delta_m(z)}{\delta_m(z=0)}$ is the linear growth factor of matter perturbations and $\sigma_8(z=0)$ is the present value of mass variance. For the GCG model, the free parameters are $\xi$ and $\sigma_8(z=0)$ and their best values at $1\sigma$, $2\sigma$, and $3\sigma$ confidence levels can be seen in Table (\ref{tab2}).

\begin{figure}[!t]
\centering
\includegraphics[width=.8\columnwidth]{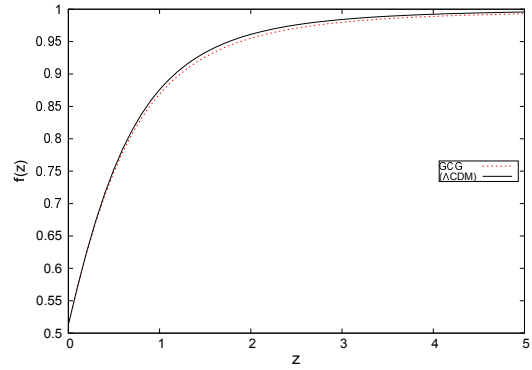}
\caption{: The evolution of matter growth function $f(z)$ vs redshift in the context of the GCG model. The values of background cosmological parameters are fixed from Table (\ref{tab1}). The values of $\xi$ and $\sigma_8(z=0)$ are fixed by using the constraints in Table (\ref{tab2}). The concordance $\Lambda$CDM is shown for comparison.}
\label{fig4}
\end{figure}

Figure (\ref{fig4}) shows the evolution of growth rate function, $f$, versus redshift. It can be seen that the amplitude of matter perturbations is reduced by the influence of dark energy at low redshifts. So, the effect of dark energy component at low redshifts in both GCG and $\Lambda$CDM models is the suppression of the growth rate of matter perturbations. It is worth noting that the effect of dark energy on the perturbations growth at high redshifts is negligible, since the fractional energy density of dark energy tends to zero at high redshifts. So the growth function of matter perturbations tends to unity at high redshift epochs.

\begin{figure}[!t]
\centering
\includegraphics[width=.8\columnwidth]{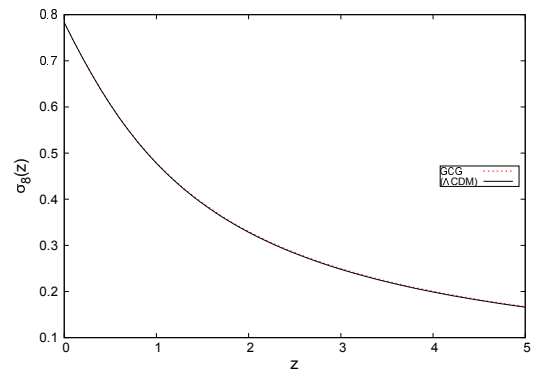}
\caption{: The evolution of mass variance $\sigma_8(z)$ vs redshift in the context of the GCG model. The background cosmological parameters and the concordance $\Lambda$CDM are the same as in Figure (\ref{fig4}).}
\label{fig5}
\end{figure}

We observe that the suppression of the amplitude of matter fluctuations in the GCG model starts almost at the same time as that in the $\Lambda$CDM model. The evolution of $\sigma_8(z)$ versus redshift is computed for the GCG and $\Lambda$CDM models in Figure (\ref{fig5}). Additionally, the perturbation variances are nearly equal $(f_{\Lambda CDM}\cong f_{GCG})$ and grow with the scale factor in both models.

Now, we calculate the theoretical value of $f(z)\sigma_8(z)$ in the context of the GCG model. Using 18 distinct and independent observational growth rate data points (\citealt{Basilakos2017}), a statistical least square analysis gives
\begin{equation}
     \chi_{gr}^2=\sum_{i=1}^{N}\frac{\left[f(z_i)\sigma_8^{(th)}(z_i)-f(z_i)\sigma_8^{(obs)}(z_i)\right]}{\sigma_i^2},
\end{equation}
where $\sigma_i$ are the corresponding uncertainties, "$obs$" denotes the observed data and "$th$" represents the theoretical prediction in the GCG model. We assume that the interaction parameter $\xi$ and mass variance $\sigma_8(z=0)$ are free parameters and constrained by the growth rate data. Other cosmological parameters are assumed to be fixed to the best-fit values in Table (\ref{tab1}). So, the statistical vector $p$ in MCMC analysis contains two free parameters ($\xi$ and $\sigma_8(z=0)$) for the GCG model and $\sigma_8(z=0)$ for the concordance $\Lambda$CDM model. Finally, the results are presented in Table (\ref{tab2}). Our numerical result shows that the $\Lambda$CDM model with the smallest $AIC$ value is still the best model in cluster level. However, the difference between GCG and $\Lambda$CDM universe is less than 3. Therefore, we can roughly say that the GCG model is fitted to growth rate data as well as $\Lambda$CDM model. It should be noted that the best-fit value of $\sigma_8(z=0)$ obtained in GCG model is lower than that of the $\Lambda$CDM model (which is approximately equal to 0.07). This result explicitly shows that the tension of $\sigma_8(z=0)$ between high redshift experiments (CMB) and large-scale observations at low redshifts in the $\Lambda$CDM model can be alleviated in GCG cosmology. This may be attributed to the interaction between dark matter and dark energy in the GCG model, while this interaction is not the case in the standard cosmology.

\begin{figure}[!t]
\centering
\begin{tabular}{@{}c@{}}
\includegraphics[width=0.8\columnwidth]{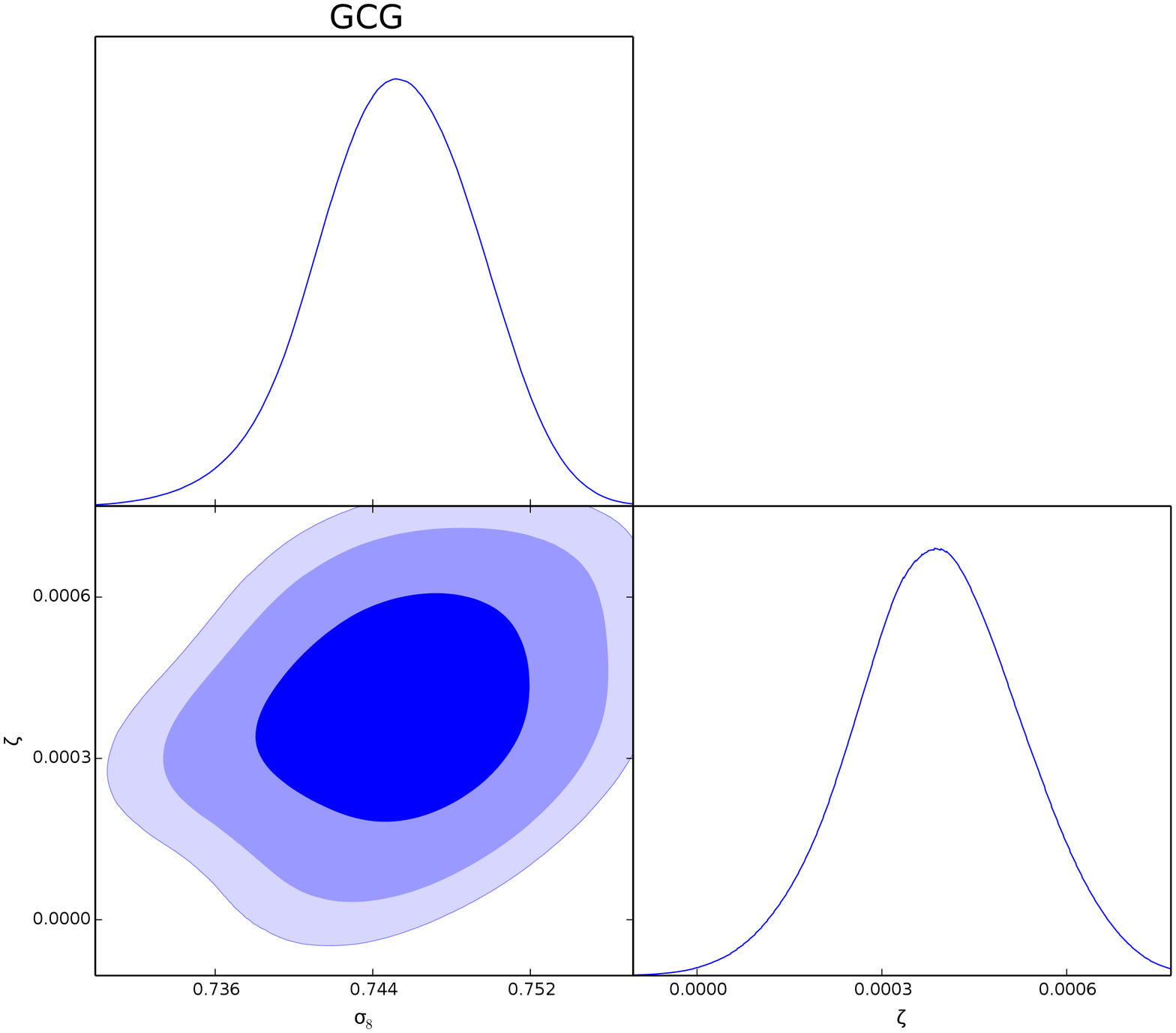}
\\
\includegraphics[width=0.8\columnwidth]{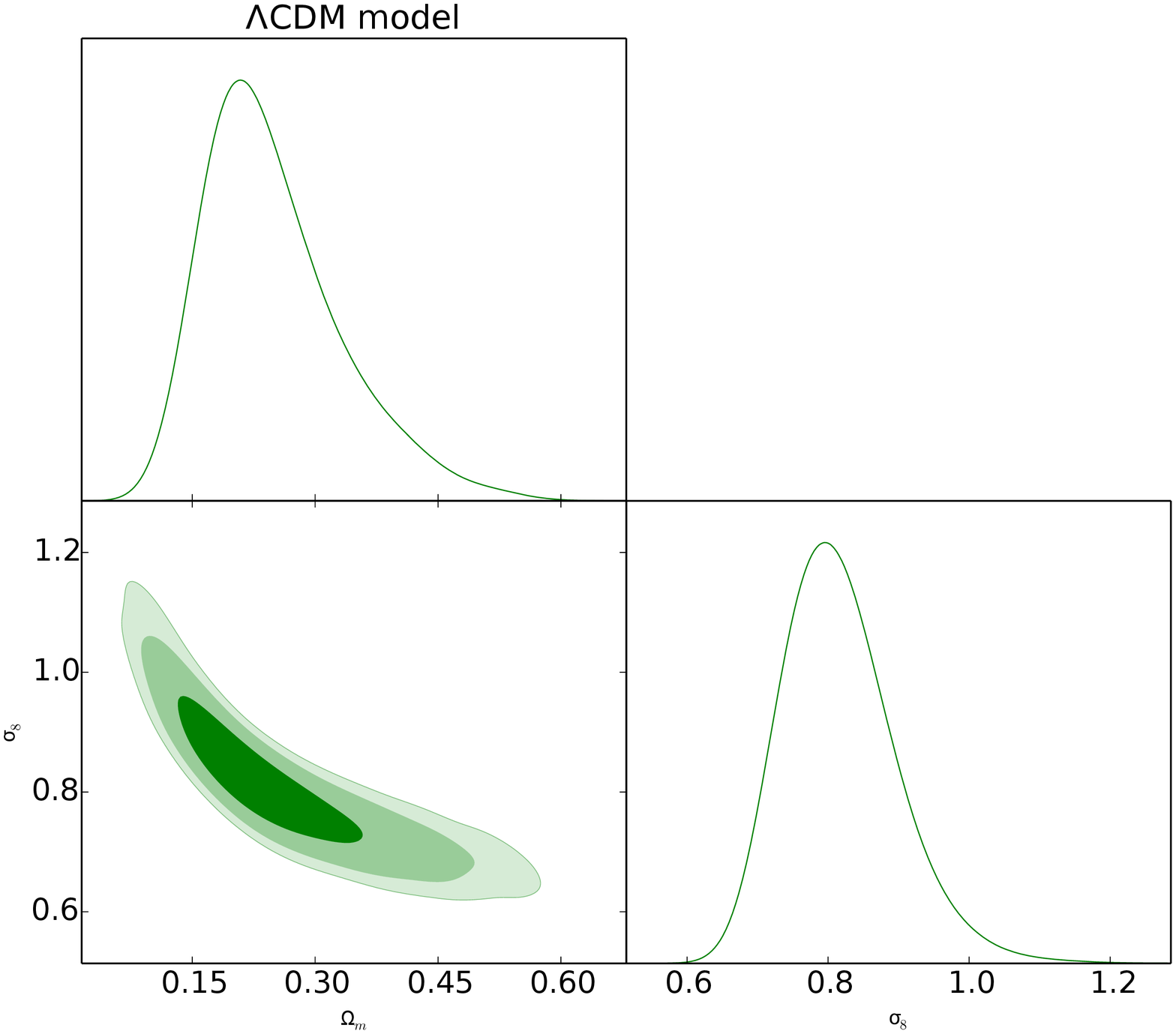}
\\
\end{tabular}
\caption{: $1\sigma$, $2\sigma$, and $3\sigma$ confidence contours and maximum likelihood functions in $\xi - \sigma_8$ and $\Omega_m - \sigma_8$ planes for GCG and $\Lambda$CDM models.}
\label{fig6}
\end{figure}

\begin{figure}[!t]
\centering
\includegraphics[width=.8\columnwidth]{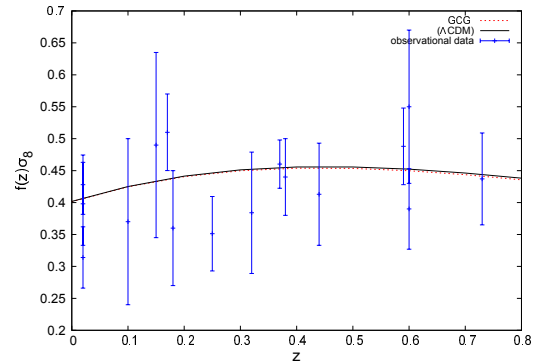}
\caption{: Predicted theoretical $f(z)\sigma_8(z)$ in GCG and $\Lambda$CDM models using the best-fit values of the cosmological parameters presented in Table (\ref{tab2}) compared with the observational growth rate data.}
\label{fig7}
\end{figure}

Figure (\ref{fig6}) demonstrates $1\sigma$, $2\sigma$ and $3\sigma$ confidence levels in the $\xi - \sigma_8$ plane, and Figure (\ref{fig7}) shows the predicted theoretical $f(z)\sigma_8(z)$ for the GCG model using the best-fit values of cosmological parameters presented in Tables (\ref{tab1}) and (\ref{tab2}). These results can be compared with the implications of Figure (\ref{fig7}), in which the predicted growth rate function $f(z)$ in the GCG model is smaller than that in the $\Lambda$CDM model, while the value of $f(z)\sigma_8(z)$ is greater in GCG than in $\Lambda$CDM. Therefore, it can be concluded that the product of growth rate function and mass variance, i.e., $f(z)\sigma_8(z)$, of GCG is consistent with that of $\Lambda$CDM (see Figure (\ref{fig7}).

\begin{table*}
 \centering
\tabularfont
 \caption{Numerical results for GCG and $\Lambda$CDM models from  statistical MCMC analysis by using cosmological growth rate data in cluster scale. The best-fit values of the cosmological parameters $\xi$ and $\sigma_8(z=0)$ are presented at their $1\sigma$, $2\sigma$, and $3\sigma$ confidence levels.}
\begin{tabular} { l  c c c c c c c  c c}
\hline		
 Parameters &GCG &$\Lambda$CDM\\
\hline
{$\xi$} &
0.00039 $ \begin{array}{c}{+0.00013+0.00027+0.00034} \\ {-0.00013-0.0027-0.0035}\end{array}$
& -
\\
 {$\sigma_8(z=0)$} &
 0.7453$ \begin{array}{c}{+0.0040+0.0079+0.0099} \\ {-0.0036-0.0081-0.0110}\end{array} $
 &0.816$ \begin{array}{c}{+0.064+0.16+0.27} \\ {-0.091-0.15-0.17}\end{array}$
\\
{$\chi_{min}^{2}$ }&
16.310 & 13.58
\\
\hline
\end{tabular}\label{tab2}
\end{table*}

\section{Results and discussion}

The present study aims at investigating the important features of the GCG model at both background and perturbation levels in three steps. In the first step, the system of the main differential equations is solved at both background and perturbation levels. The behavior of the basic cosmological quantities ($\Omega(z)$, $w(z)$, $E(z)$) is then investigated to determine the general characteristics of the GCG model (see Figures (\ref{fig1}) and (\ref{fig3})). 
It is shown that the EoS remains in the quintessence regime within the range --$1 \leq w_{GCG} \leq -\frac{2}{3}$. According to Figure (\ref{fig3}), the GCG model behaves similarly to non-relativistic dark matter ($w=0$) at early times ($z\rightarrow \infty$) and to dark energy at late times. Moreover, $E_{GCG}(z)>E_{\Lambda CDM}(z)$ indicates a faster expansion of GCG than the standard model. As expected, in the GCG model, the energy of the universe is currently dominated by the dark energy component. In other words, the dark energy density begins to grow at $z \sim 1$ and dominates the dark matter density at $z \sim 0.2$ (Figure (\ref{fig1})). As a result, $f(z)$ tends to unity in both GCG and $\Lambda$CDM models at large redshifts. Additionally, the decrease in $f(z)$ for small redshifts is due to the effect of dark energy (Figure (\ref{fig4})).

In the second step, a joint statistical analysis is performed exploiting the latest geometrical and growth rate data. We find that the joint statistical analysis in the context of flat FRW Universe can place tight constraints on the main cosmological parameters. While $\eta=\alpha+1$ in the GCG model, we find $\alpha>0$, which means energy transfer from dark matter to dark energy in the this model. Our results at $1\sigma$ confidence were compared with some previous works in Table \ref{tab3}. The values of $A_s$ in the $2nd$, $3rd$, $4th$, and $5th$ rows are consistent. Since it is important that $\alpha$ is greater or less than zero, the values of $\alpha$ obtained in this work are close to those in the $2nd$ and $5th$ rows of Table \ref{tab3}. We show that the present values of mass variance, $\sigma_8(z=0)$, obtained in the GCG model is less than that of the standard $\Lambda$CDM cosmology. This indicates that the big tension of $\sigma_8$  between high and low redshift observations appeared in the standard cosmology can be alleviated in the GCG model.

In the third step, we obtain $AIC_{\Lambda CDM}=577.9$ and $AIC_{GCG} = 600.6$ for the observational data at expansion level. The large value of $\Delta AIC=22.7$ indicates a better agreement between the standard $\Lambda$CDM cosmology and the cosmological observations at the expansion level. Therefore, according to our analysis, considering observational data in the background cosmology, the simple $\Lambda$CDM model is still the best model. It is worth noting that the literature on the dynamical dark energy models has presented the same results (for example, see (\citealt{Malekjani2018}; \citealt{Rezaei2019})).

\begin{table*}
 \centering
 \tabularfont
 \caption{Constrained values of $A_s$ and $\alpha$ at $1\sigma$ confidence region obtained from our analysis in this study and also some previous works.}
\begin{tabular} { l  c  c c c c c c c  c c c}
\hline		
 Row & Reference & {$A_s$} & {$\alpha$} & Data
\\
\hline
1 & {\tiny(\citealt{Hang2019})} & 0.642$ \begin{array}{c}{+0.097} \\ {-0.093}\end{array}$ & -0.1688$ \begin{array}{c}{+0.1456} \\ {-0.2143}\end{array}$ & CMB$+$CC
\\
2 & {\tiny(\citealt{Hang2019})} & 0.730$ \begin{array}{c}{+0.047} \\ {-0.047}\end{array}$ & 0.0181$ \begin{array}{c}{+0.1029} \\ {-0.2129}\end{array}$ & CMB$+$JLA
\\
3 & {\tiny(\citealt{Hang2019})} & 0.727$ \begin{array}{c}{+0.040} \\ {-0.039}\end{array}$ & -0.0156$ \begin{array}{c}{+0.982} \\ {-0.1380}\end{array}$ & CMB$+$JLA$+$CC
\\
4 & {\tiny(\citealt{Liang2011})} & 0.7475$ \begin{array}{c}{+0.0556} \\ {-0.0539}\end{array}$ & -0.0250$ \begin{array}{c}{+0.1760} \\ {-0.1326}\end{array}$ & CMB$+$BAO$+$SNIa
\\
5 & {\tiny(\citealt{Malekjani2011})} & 0.76$ \begin{array}{c}{+0.026} \\ {-0.039}\end{array}$ & 0.033$ \begin{array}{c}{+0.066} \\ {-0.0071}\end{array}$ & CMB$+$BAO$+$SNIa$+$H(z)$+$BBN
\\
6 & {\tiny This work} & 0.774$ \begin{array}{c}{+0.022} \\ {-0.022}\end{array}$ & 0.096$ \begin{array}{c}{+0.059} \\ {-0.074}\end{array}$ & CMB$+$BAO$+$SNIa$+$H(z)$+$BBN
\\
\hline
\end{tabular}\label{tab3}
\end{table*}

\section{Conclusion}
In this work, the cosmological properties of the GCG model was studied and the results of different combinations of observational data points were compared. Moreover, for describing the evolution of the cosmic fluid, a comparison was made between the results of the GCG and $\Lambda$CDM model. The growth rate of matter perturbations for GCG cosmology was investigated. It was shown that in this model, the amplitude of matter perturbations can be suppressed by the dark energy component, as done by the $\Lambda$ sector in the standard $\Lambda$CDM model. We put  observational constraints on the background evolution of the GCG cosmology based on data from SNIa (Union2.1 sample), CMB, BAO, BBN, and Hubble data. Our numerical results based on the statistical analysis indicate that the standard $\Lambda$CDM model is still a better model compare to GCG model. However, the current expansion rate of the Universe in GCG model is in good agreement with that of the standard model. It was also shown that the GCG parametrization is almost consistent with the growth rate data in cluster scale as much as the concordance $\Lambda$CDM model, while the tension of $\sigma_8$ can be alleviated in the GCG model.
\balance
\vspace{-1em}



\end{document}